\documentstyle[prd,aps,twocolumn,psfig]{revtex}

\begin{document}

\def\lsim{\
  \lower-1.2pt\vbox{\hbox{\rlap{$<$}\lower5pt\vbox{\hbox{$\sim$}}}}\ }
\def\gsim{\
  \lower-1.2pt\vbox{\hbox{\rlap{$>$}\lower5pt\vbox{\hbox{$\sim$}}}}\ }

\title{Recycling the universe using scalar fields}
\author {Nissim Kanekar$^{1,\star}$, Varun Sahni $^{2,\dagger}$ and
Yuri Shtanov $^{3,\ddagger}$}
\address{$^1$National Centre for Radio Astrophysics, Pune 411007, India}
\address{$^2$Inter-University Centre for Astronomy \& Astrophysics,
Post Bag 4, Pune 411007, India}
\address{$^3$Bogolyubov Institute for Theoretical Physics, Kiev 03143,
Ukraine}

\maketitle

\bigskip

\begin{abstract}
We examine the behaviour of a closed oscillating universe filled with a
homogeneous scalar field and find that, contrary to naive expectations, such
a universe expands to larger volumes during successive expansion epochs.
This intriguing behaviour introduces an arrow of time in a system which is
time-reversible. The increase in the maximum size of the universe is closely
related to the work done on/by the scalar field during one complete
oscillatory cycle which, in turn, is related to the asymmetry in the scalar
field equation of state during expansion and collapse. Our analysis shows
that scalar fields with polynomial potentials $V(\phi) = \lambda \phi^q$, $q
> 1$ lead to a growing oscillation amplitude for the universe: the increase
in amplitude between successive oscillations is more significant for smaller
values of $q$. Such behaviour allows for the effective recycling of the
universe. A recycled universe can be quite old and can resolve the flatness
problem. These results have strong bearing on cosmological models in which
the role of dark matter is played by a scalar field. They are also relevant
for chaotic inflationary models of the early universe since they demonstrate
that, even if the universe fails to inflate the first time around, it will
eventually do so during future oscillatory cycles. Thus, the space of
initial conditions favourable for chaotic inflation increases significantly.
\end{abstract}

\section{Introduction}
The idea of a cyclical oscillating universe---one that is continuously
reborn from the ashes of a previous existence---finds expression in the
philosophical and cultural beliefs of many ancient civilizations
\cite{eliade,jaki,munitz,zimmer}.

Within the framework of modern relativistic cosmology, oscillating models
(analytically continued through the big bang singularity) arise naturally as
exact solutions of the Einstein field equations for a spatially closed
universe consisting of a perfect fluid. Since all expansion-contraction
cycles in such models are identical, one might feel that an oscillating
universe containing an infinite number of cycles would be infinitely old,
somewhat resembling, on the average, a steady-state model. Dissipative
processes leading to entropy growth, however, change this picture radically.
As originally demonstrated by Tolman \cite{tolman}, the growth of entropy
increases the total volume of the universe at the maximum of each expansion
cycle; this observation has several important consequences, some of which
are summarised below.

(i) Tolman strongly felt that the possibility of thermodynamically recycling
the universe would have a ``liberalizing action on our general thermodynamic
thinking'' since it would dispel the notion that ``the principles of
thermodynamics necessarily require a universe which was created at a finite
time in the past and which is fated for stagnation and death in the future''
\cite{tolman}. Thus, the oscillating universe was seen to present a credible
alternative to the idea of the thermodynamic heat death postulated by
nineteenth century physicists and popular in this century as well. (The
latter may still be possible in a flat/open universe.)

(ii) As demonstrated by Zeldovich and Novikov \cite{zn83}, the increase in
entropy from cycle to cycle suggests that an oscillating universe could not
have had an infinitely long total duration since, given the present (finite)
value of its total entropy and postulating that the entropy increase from
cycle to cycle is finite, one is led to conclude that the number of cycles
preceeding the present one is also finite. An oscillating universe cannot
therefore be infinitely old and must have been created, perhaps quantum
mechanically, at some point in the past. Since the total mass (energy) of a
closed universe is zero, its creation from the vacuum does not violate any
known laws of conservation and is therefore possible, in principle
\cite{fomin75}.

(iii) An important consequence of an oscillating universe with an increasing
expansion maximum at every cycle, is that the horizon and flatness problems
are gradually ameliorated as the universe grows older, larger, and flatter
during each successive expansion cycle. The oscillating universe may thus
present a credible alternative to the inflationary scenario in this respect.

An oscillating universe can also have other important cosmological
implications. For instance, relics of an earlier expansion epoch may be
measurable today \cite{rees}. Starobinsky, in his seminal paper on graviton
production in an inflationary universe \cite{star79}, showed that gravity
waves would also be created in an oscillatory universe; further, he derived
an expression for their amplitude and spectrum which allows us to infer
details of previous expansion epochs from measurements made today.

Finally, the idea of an oscillating universe also appears in the
quasi-steady-state cosmology of Hoyle, Burbidge, and Narlikar
\cite{hoyle1,hoyle2}, in which alternate cycles of expansion and contraction
modulate an exponentially expanding background geometry, with the creation
of matter being most intense at the minimum of each cycle.

A key issue, which remains at present unresolved, relates to physical
mechanisms which can cause the universe to bounce at the end of each cycle.
The singularity theorems of Penrose and Hawking suggest that a singular
state necessarily arises at the end of contraction in a closed
Friedman--Robertson--Walker (FRW) universe described by general relativity,
if matter satisfies certain `energy conditions' \cite{haw}. However, it is
not clear whether, at very high curvatures ${\cal R} \sim l_{\mathrm
P}^{-2}$, matter would behave as it does under `normal' conditions when the
space-time curvature is much smaller than the Planck value, ${\cal R} \ll
l_{\mathrm P}^{-2}$. Furthermore, at ${\cal R} \sim l_{\mathrm P}^{-2}$,
particle production and vacuum polarisation effects are likely to be
significant; the resulting vacuum expectation value of the energy-momentum
tensor $\langle T_{ik}\rangle$ need not satisfy the above energy conditions,
due to which the space-time metric could very well `bounce' without ever
reaching a singular state \cite{zn83,bd82}. Finally, we should mention that
a `bouncing' universe may also arise in quantum cosmology or as a
consequence of the duality conditions which are generic features of
superstring-inspired cosmological models. The limiting curvature hypothesis
of Markov \cite{markov} (see also \cite{mukhanov}) might also lead to a time
symmetric bounce. Other physical mechanisms leading to singularity avoidance
are discussed in \cite{linde,barrow}.

\begin{figure}
\centering \psfig{file=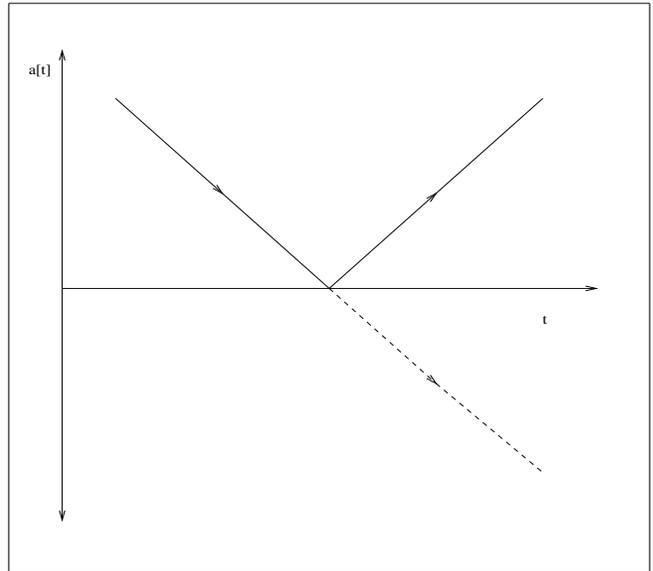,width=3.37truein,height=3.0truein,angle=-90}
\caption{The `bounce' at $a = 0$, $t = 0$ (solid line) is replaced by a
continuous transition of the scale factor through $a = 0$ towards negative
values of $a$ (dotted line). Both representations are physically equivalent
since only the {\em absolute\/} value of $a$ has physical significance.}
\label{fig1}
\end{figure}

The bouncing of the universe in the vicinity of the singularity, $a = 0$, is
perceived less dramatically by extending the range of the scale factor to
negative values. Since only the absolute value of the scale factor, $|a|$,
has direct physical meaning, such an extension is quite legitimate (see, for
instance, \cite{kuz} and Fig.~\ref{fig1}).  The notion of a `bounce' is then
replaced by the more natural notion of a continuous `passage' of the scale
factor through the value of $a = 0$ to the region of negative values $a <
0$.  This also can be viewed as the change of the (unphysical?) spatial
orientation of the universe after its passage through zero value of the
scale factor \cite{fomin}. We will use this notion in our formulation of
`natural' conditions at the bounce.  Of course, the dynamics of the universe
in the close vicinity of the point $a = 0$ is beyond classical description,
even in this model.

This paper presents a radical departure from most previous work on
oscillating models in which entropy production associated with dissipative
processes led to the growth of the expansion maximum of each cycle. Our
results show that an alternate mechanism which does not require entropy
production exists, also leading to increasing oscillatory cycles.  We
demonstrate that the presence of a massive scalar field (in a closed FRW
universe), under certain reasonable conditions at the bounce, gives rise to
growing expansion cycles, the increase in expansion amplitude being related
to the work done by/on the scalar field during the expansion/contraction of
the universe. (The presence of other matter fields, in addition to the
scalar field, does not affect our conclusions, as long as interactions
between such fields and the scalar are sufficiently weak.) Our results have
important consequences for the inflationary universe scenario, one of the
standard paradigms of cosmology.

\section{Scalar fields in a closed universe.}

\begin{figure}
\centering \psfig{file=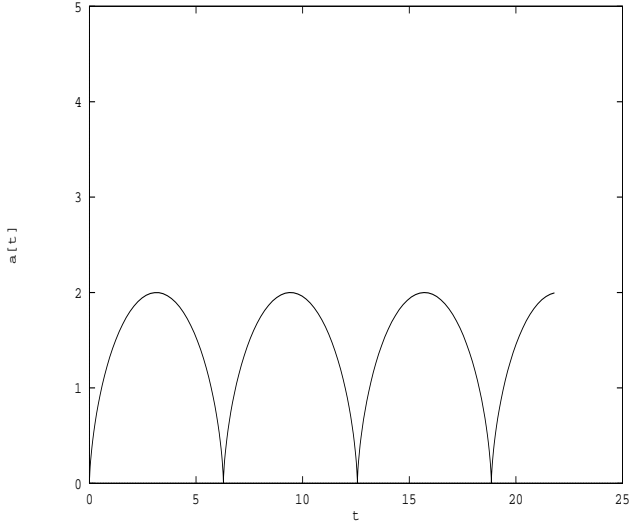,width=3.37truein,height=3.0truein,angle=-90}
\caption{The cycloid solution for a matter dominated universe. Note that the
amplitude of the cycloid does not increase from one cycle to the next.}
\label{fig2}
\end{figure}

The recent detection of anisotropy in the cosmic microwave background on
degree scales appears to favour a closed FRW universe with $\Omega_{\mathrm
total} = 1.11 \pm 0.07$ \cite{boom_max}. A spatially closed space-time is
described by the metric
\begin{equation}
ds^2 = dt^2 - a^2 (t) \left( {dr^2 \over 1 - r^2} + r^2 d \theta^2 + r^2
\sin^2 \theta d \phi^2 \right) \, , \label{eq:1}
\end{equation}
where $a(t)$ is the cosmic expansion factor; units in which the
speed of light $c = 1$ are used throughout this paper. In a closed universe,
the Einstein equations acquire the well-known form
\begin{equation}
\left( {\dot a \over a} \right)^{2} = {8 \pi G \over 3} \rho - {1 \over a^2}
\, ,  \label{eq:2}
\end{equation}
\begin{equation}
{\ddot{a} \over a} = - {4 \pi G \over 3} (\rho + 3 P ) \, , \label{eq:3}
\end{equation}
where $\rho$ and $P$ are, respectively, the total energy density and
pressure of the various components of the universe. For a
radiation-dominated universe, we have $P = \rho / 3$, and the field
equations can be solved exactly in terms of the conformal time coordinate
$\eta = \int dt/a(t)$, to yield the solution
\begin{equation}
a(\eta) = A \sin \eta \, , \qquad t = A (1 - \cos \eta) \, ,
\label{eq:cycloid}
\end{equation}
which describes a semicircle in the $a$--$t$ plane. The solution for a
matter dominated universe (with $P = 0$) is
\begin{equation}
a(\eta) = A(1 - \cos \eta) \, , \qquad t = A(\eta -\sin \eta) \, ,
\label{eq:cycloid1}
\end{equation}
which describes a cycloid.  After solution (\ref{eq:cycloid}) is extended
periodically in time, both (\ref{eq:cycloid}) and (\ref{eq:cycloid1})
describe a periodic evolution with an infinite number of identical
expansion--contraction cycles, as shown in Fig.~\ref{fig2}.

The presence of a bulk viscosity $\zeta$ results in a modification of the
cosmological fluid pressure $P$ to
\begin{equation}
P = P_0 - 3\zeta H \, . \label{eq:4}
\end{equation}
Here, $P_0$ is the equilibrium pressure and $H = {\dot{a}}/{a}$ is the
Hubble parameter. (An example of bulk viscosity is provided by a fluid in
which energy is easily exchanged between translational and
rotational/internal degrees of freedom, an example being a gas of rough
spheres \cite{wein}.) From Eq.~(\ref{eq:4}) one finds that, during
expansion, $H > 0$ and $P < P_0$, whereas during collapse, $H < 0$ and $P >
P_0$. This asymmetry during the expanding and contracting phases results in
the growth of both energy and entropy, in the words of Tolman \cite{tolman},
``if the pressure tends to be greater during a compression than during a
previous expansion, as would be expected with a lag behind equilibrium
conditions, an element of fluid can return to its original volume with
increased energy and hence also with increased entropy.'' The increase in
energy makes the amplitude of successive expansion cycles larger enabling
the universe to spend ``a greater and greater proportion of its period in a
condition of lower density $\ldots$ even though a return to higher densities
would always occur.''

Although Tolman linked the asymmetry in pressure during expansion and
collapse to the production of entropy, we will show that such an asymmetry
also arises for non-dissipative Lagrangians such as those describing a
massive scalar field in an FRW space-time. In this case, the asymmetry in
pressure leads to a significant increase in the energy of the scalar field
and results in an increase of the maximum volume of the oscillating
universe.

The Lagrangian density for a scalar field has the form
\begin{equation}
{\cal L} = \frac12 g^{ij} \partial_i \phi \partial_j \phi - V(\phi) \, .
\label{eq:lagr}
\end{equation}
The 0-0 Einstein equation for a homogeneous scalar field in the closed FRW
universe (\ref{eq:1}) becomes
\begin{equation}
H^2 = \frac{8 \pi G}{3} \left[ \frac12 \dot{\phi}^2 + V (\phi) \right] -
\frac{1}{a^2} \, , \label{eq:5}
\end{equation}
and the energy density and pressure of the scalar field are, respectively,
\begin{equation}
\rho_\phi \simeq \frac12 \dot{\phi}^2 + V (\phi) \, , \qquad  P_\phi \simeq
\frac12 \dot{\phi}^2 - V (\phi) \, . \label{eq:6}
\end{equation}
The scalar-field equation of motion is
\begin{equation}
\ddot \phi + 3 H \dot \phi + \frac{dV}{d\phi} = 0 \, . \label{eq:7}
\end{equation}
The motion of $\phi$ arises in response to the dual action of the
accelerating force $dV / d \phi$ and the damping term $ 3 H \dot \phi$. In
the chaotic inflationary scenario with $V \propto \phi^n$ ($n = 2, 4$), for
sufficiently large values of $\phi$, namely, for $\phi \gsim m_{\mathrm P}$,
where $m_{\mathrm P}$ is the Planck mass, the damping caused by the
expansion of the universe ($H > 0$) settles the scalar field into a
`slow-roll' regime during which $\frac12 \dot{\phi}^2 \ll V$ and $P \simeq -
\rho$. Since the equation of state of the scalar field mimics that of the
cosmological constant, the expansion of the universe is inflationary, $a
\propto \exp \int H(t) dt$. Exactly the reverse situation arises when the
universe contracts. In this case, we have $H < 0$, and the term $3 H \dot
\phi$ now accelerates the motion of $\phi$ instead of damping it, as it did
during expansion. As a result, the kinetic energy of the scalar field
becomes much larger than its potential energy, $\frac12 \dot{\phi}^2 \gg V$,
and the resulting equation of state becomes $P = \rho$ (sometimes called the
equation of state of `stiff' matter). Consequently, the scalar field in a
closed universe satisfies two generic regimes \cite{zel}
\begin{equation} 
P \simeq - \rho \ \ \mbox{for} \ \ H > 0 \ \ \mbox{(expansion)} \, , \label{eq:8a}
\end{equation}
\begin{equation}
P \simeq \rho \ \ \mbox{for} \ \ H < 0 \ \
\mbox{(contraction)} \, . \label{eq:8}
\end{equation}
For polynomial potentials, $V \propto \phi^n$, these two regimes are
separated by an epoch during which the scalar field oscillates about its
minimum value while its equation of state mimics that of dust: $\langle P
\rangle = 0$ ($n = 2$), or radiation: $\langle P \rangle = \langle \rho
\rangle / 3$ ($n = 4$). (The time average is taken over many oscillations of
the field.) It is interesting that the quantity $P / \rho$, when plotted as
a function of the expansion factor, resembles a hysteresis curve. The area
enclosed by the curve is related to the work done by/on the scalar field
during the expansion of the universe $\delta W = \oint P dV$.

If one postulates that the universe `bounces' during contraction, then one
can expect the `work done' $\delta W$ during a given expansion cycle to be
converted into `expansion energy', resulting in the growth in amplitude of
each successive expansion cycle. An estimate of the increase in the
expansion maximum can be obtained from the following elementary
considerations: setting $H = 0$ in (\ref{eq:2}) we obtain
\begin{equation}
\rho_* = \frac{3}{8\pi G a_{\mathrm max}^2} \, , \label{eq:9}
\end{equation}
where $\rho_*$ is the density of the universe at the expansion maximum.
Since the total mass of a closed universe is $M = 2 \pi^2 a^3 \rho$, this
gives
\begin{equation}
M = \frac{3\pi}{4 G}a_{\mathrm max} \, .
\end{equation}
Equating the increase in energy, $\delta E = \delta M$, to the work done
during a single expansion-contraction cycle, $\delta E = \delta W$, we
obtain
\begin{equation}
\Delta a_{\mathrm max} = \frac{4G}{3\pi} \oint P dV \, , \label{eq:10}
\end{equation}
which relates the increase in the expansion maximum between two successive
cycles $\Delta a_{\mathrm max}$ to the work done $\delta W = \oint P dV$.
(The increase in energy $\delta E$ is brought about at the expense of the
gravitational field energy, since the total energy of a closed universe is
identically zero \cite{tolman,ll,zn83}.)

We probe further implications of this analysis by studying equations
(\ref{eq:5})--(\ref{eq:7}) numerically for the class of chaotic potentials
$V (\phi) = \lambda \phi^q$, $q > 0$, and $V (\phi) = V_0\, \exp (- \mu
\phi)$. The potential $V (\phi) \propto \phi^q$ has been discussed both in
the context of inflation \cite{linde90} and as a candidate for cold dark
matter \cite{peebles}. Exponential-based classes of potentials are known to
have important cosmological consequences both within the inflationary
framework \cite{lucchin} and as candidates for quintessence and cold dark
matter \cite{ss2000}. (An early use of the exponential potential
in the context of a closed universe may be found in \cite{sfs92}.)
Before describing our results, it is necessary to first
discuss the type of `bouncing' conditions which are imposed at the start of
each expansion-contraction cycle. As remarked earlier, it is likely that a
unified theory of the weak, strong, and gravitational interactions will
provide a key to understanding physical processes which operate during the
strong-curvature regime of a contraction cycle. In the absence of such a
theory, we will do the next best thing and set conditions which we feel are
both simple and natural and may therefore arise in a future `physically
complete' theory of the early universe.

The conditions imposed at the bounce are, $a \to a$, $\dot a \to {} - \dot
a$, $\phi \to \phi$, and $\dot \phi \to \dot \phi$, (the bounce is assumed
to occur instantaneously). Thus, the effect of the bounce is to reverse the
direction of motion of the universe (since ${\dot a} \to {} - \dot a$), with
all other quantities remaining unchanged. In the introduction, we mentioned
the possibility of extending the range of the scale factor to negative
values and then replacing the idea of a `bounce' by the somewhat more
appealing idea of `passage' to negative values of the scale factor. This
viewpoint leads to the following reasonable conditions at the bounce: $a \to
{}- a$, $\dot a \to \dot a$, $\phi \to \phi$, $\dot \phi \to \dot \phi$,
which is equivalent to the `conventional' prescription $a \to a$, $\dot a
\to {}- \dot a$, $\phi \to \phi$, $\dot \phi \to \dot \phi$ adopted in this
paper. Further, we assume that the bounce arises at scales at which quantum
gravitational effects become important, i.e., at Planck scales. Two
possibilities then exist for the {\it location\/} of the bounce:

(i) The bounce occurs when the {\it curvature\/} of the universe becomes
comparable to the Planck scale, in other words, the bounce takes place at a
fixed value $a_{\mathrm min} \simeq l_{\mathrm P}$ of the scale factor.

(ii) The bounce occurs when the {\it energy density\/} in the scalar field
crosses the Planck energy, i.e., when $\frac12 \dot{\phi}^2 + V (\phi)
\simeq m_{\mathrm P}^4$.

\begin{figure}
\centering \psfig{file=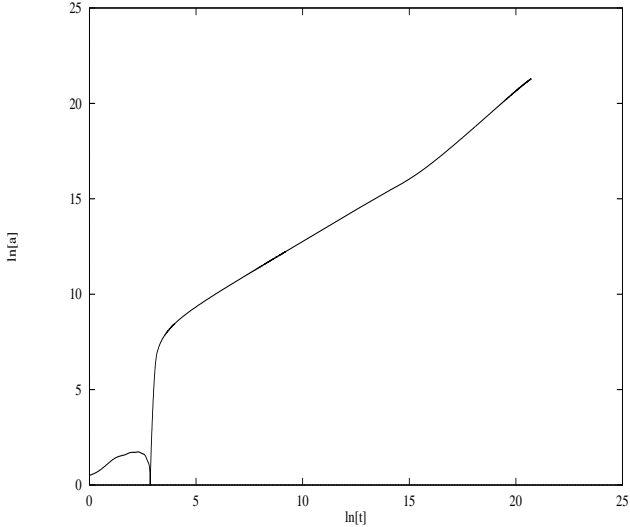,width=3.37truein,height=3.0truein,angle=-90}
\caption{The expansion factor for an oscillating universe satisfying the
boundary condition (i).} \label{fig3a}
\end{figure}

\begin{figure}
\centering \psfig{file=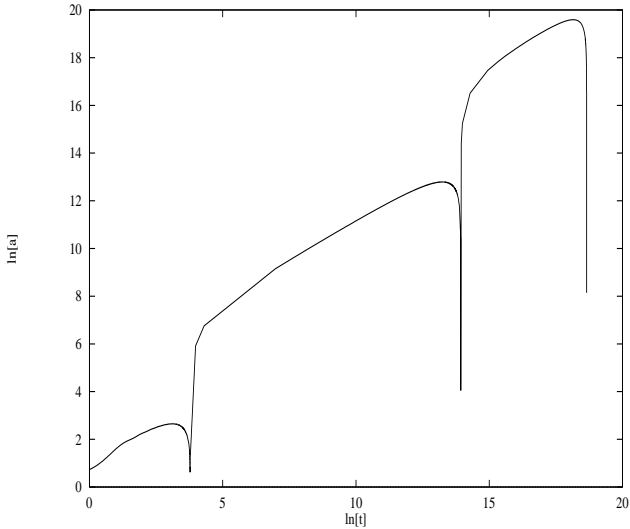,width=3.37truein,height=3.0truein,angle=-90}
\caption{The expansion factor for an oscillating universe satisfying the
boundary condition (ii). Note the monotonic increase in successive expansion
maxima (plotted on the logarithmic scale).} \label{fig4a}
\end{figure}

Both the above possibilities are considered in the current work, with
results shown in Figs. \ref{fig3a} and \ref{fig4a}, respectively, for the
boundary conditions (i) and (ii), and for the potential $V (\phi) =
\frac{1}{2} m^2 \phi^2$. In Fig.~\ref{fig3a}, we follow the expansion factor
through one expansion-contraction cycle; it can be seen that the universe
inflates at the beginning of the second cycle, resulting in a very large
universe and a correspondingly large value of $a_{\mathrm max}$.

Next, in implementing condition (ii), we will work with what we feel is the
`worst case scenario' for inflation, namely, we set
\begin{equation}
\frac12 \left. \dot \phi^2 \right|_{a = a_{\mathrm min}} \simeq m_{\mathrm
P}^4 \gg \left. V(\phi) \right|_{a = a_{\mathrm min}} \, , \label{eq:init1}
\end{equation}
\begin{equation}
\left. \phi \right|_{a = a_{\mathrm min}} < \phi_* \simeq m_{\mathrm P}
\sqrt{\Omega / 6 \pi} \label{eq:init}
\end{equation}
at the start of the first expansion cycle. Note that $V (\phi) = \frac{1}{2}
m^2 \phi^2$ and $\Omega = 8 \pi G \rho_\phi / 3 H^2 > 1$.

The initial conditions (\ref{eq:init1}) and (\ref{eq:init}) guarantee that the condition for
accelerated expansion $\dot \phi^2 < V (\phi)$ is not met and thus ensure
that the inflationary regime {\em does not commence\/}. Our results, shown
in Fig.~\ref{fig4a}, are interesting; we find that, although the universe is
prevented from inflating during the first cycle, its amplitude during
subsequent cycles increases, resulting finally in inflation! The physical
reasons for a growing oscillation amplitude are simple to understand: the
large kinetic energy of the scalar field during the collapse phase of a
cycle drives $\phi$ to regions higher up on the potential at the
commencement of the next expansion cycle, until the field finally reaches an
amplitude $\phi > \phi_*$, which is large enough to make the universe
inflate.

\begin{figure}
\centering \psfig{file=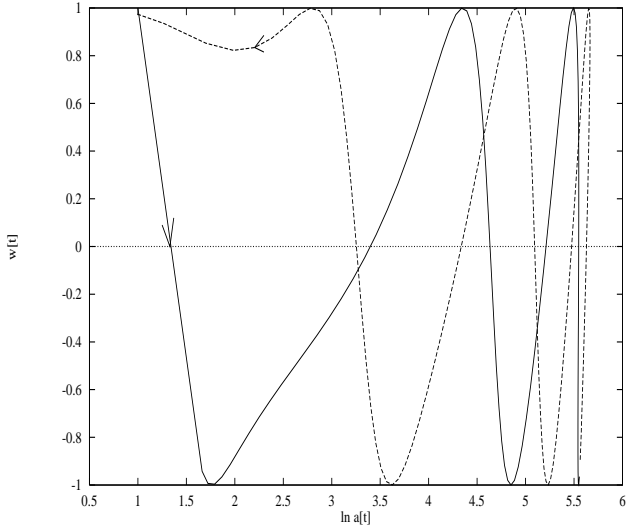,width=3.37truein,height=3.0truein,angle=-90}
\caption{The scalar-field equation of state $\omega = P / \rho$ is shown for
the first cycle of Fig.~\ref{fig3a}. Solid/broken lines correspond  to the
expanding/collapsing epochs. Note that during the pre-oscillation and
post-oscillation epochs, the behaviour of $P / \rho$ resembles a hysteresis
curve, which contributes significantly to the total `work done' $\delta W =
\oint P dV$.} \label{fig3b}
\end{figure}

\begin{figure}
\centering \psfig{file=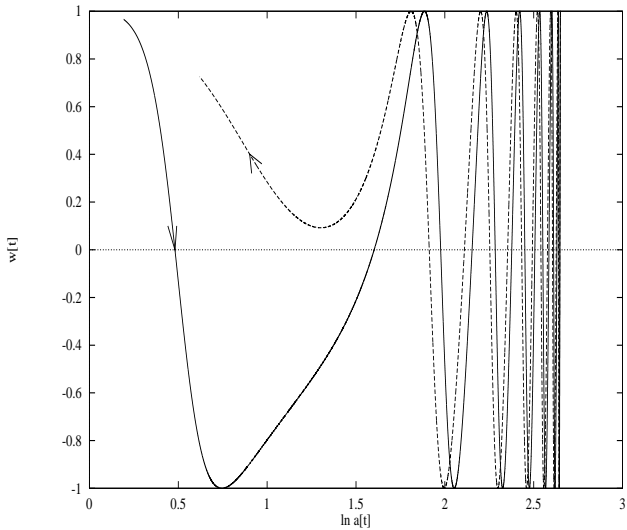,width=3.37truein,height=3.0truein,angle=-90}
\caption{The hysteresis curve corresponding to the first cycle of
Fig.~\ref{fig4a}.} \label{fig4b}
\end{figure}

The hysteresis curves corresponding to the first oscillation of the
expansion factor in Figs. \ref{fig3a} and \ref{fig4a} are shown in Figs.
\ref{fig3b} and \ref{fig4b}, respectively. Suprisingly, we find that the
phenomenon of `hysteresis' is present even when the conditions (\ref{eq:8a}) 
and (\ref{eq:8}) are no longer met. The resulting increase in oscillation amplitude
 emphasises the fact that, although the universe may not inflate during
its `first attempt', it will eventually do so during subsequent cycles! This
result considerably enhances the scope and appeal of inflationary models by
showing that inflation could begin from a much broader set of initial data,
provided the universe bounces when it reaches a high density. (It should be
pointed out that, for scalar fields originating from larger initial values
$\phi|_{a = a_{\mathrm min}} > \phi_*$, there will be greater hysteresis in
the equation of state and a larger associated growth in the volume of the
universe at maximum expansion.)

%

We thus find that a growth in the maximum expansion amplitude can be
achieved without any recourse to an entropy generating mechanism. In fact,
the field equations (\ref{eq:5})--(\ref{eq:7}) are non-dissipative and
time-reversible; moreover, the bouncing conditions imposed are also
time-reversible, in the sense that the time-reversed evolution will respect
the same bouncing conditions. One therefore arrives at the following
important conclusion: {\em time-asymmetry in the evolution of a closed
universe can be achieved even with time-reversible field equations and
bouncing conditions!} This result may appear counter-intuitive, especially
when compared with the behaviour of the periodic solutions
(\ref{eq:cycloid}) and (\ref{eq:cycloid1}), which are exactly time-symmetric
with respect to the expansion maximum of each cycle. The reason behind the
growing amplitude of successive expansion cycles has to do with the fact
that convex potentials (which have so far been examined) have a well defined
minimum about which the scalar field oscillates. Rapid oscillations of
$\phi$ with frequency $m \equiv d^2V/d\phi^2 \gg H$ cause the field to `mix'
in its phase-space $\lbrace \phi, {\dot\phi}\rbrace$
so that, at the time of recollapse, its location in
$\lbrace \phi, {\dot\phi}\rbrace$ is almost completely
uncorrelated with its location in $\lbrace \phi, {\dot\phi}\rbrace$
before the onset of oscillations. As a
result, the chance that the scalar field will roll up its potential during
recollapse along exactly the same trajectory down which it rolled during
expansion is essentially zero. This accounts for the fact that the equation
of state during contraction is $P = \rho$ and not $P = -\rho$, which it
would have been for exactly time-symmetric expansion--collapse. The
time-asymmetric behaviour of $\phi$ is clearly shown in Figs. \ref{fig3b}
and \ref{fig4b}, in which the solid line showing the equation of state
$P/\rho$ during expansion is displaced with respect to the broken line
showing $P/\rho$ during collapse. For exact time reversal, no such
difference between expansion and collapse would have been present; the solid
and broken lines would overlap, and the `hysteresis' seen in the figures
would not occur.
The global time-asymmetry of the evolution of an
oscillatory universe in our model is also a direct consequence of
the time-asymmetric physical conditions at the bounce,
in which the sign of $\dot a$ is reversed while the sign of
$\dot\phi$ is not. It should
be
noted at this point that, had we adopted the bouncing condition $\dot \phi
\to {} - \dot \phi$ for the scalar-field velocity, then every two
subsequent
cycles would obviously be time-symmetric with respect to the corresponding
bouncing point, and the cosmological evolution as a whole would thus be
periodic with period comprising two subsequent cycles.

The above argument is supported by our analysis of exponential potentials $V
= V_0\exp{(-\mu\phi)}$ for which we find no hysteresis in the equation of
state $P/\rho$ and no increase in the amplitude of the expansion factor
through successive oscillations. The example of an exponential
potential is easy to analyse because the evolution of the scalar field
proceeds with time towards large positive values of $\phi$ and, at such
values of the scalar field, the potential approaches an identical zero---a
trivial case for which the evolution is periodic and hysteresis is obviously
absent. Interestingly enough, hysteresis reappears when we modify the
potential to $V = V_0\, [\cosh (\lambda \phi) - 1]$, which has been proposed
as a candidate for cold dark matter in \cite{sahni_wang}. This potential has
the following limiting forms: $V \propto \exp (- \lambda \phi)$ for
$|\lambda \phi| \gg 1$, $\phi < 0$; and $V \propto \lambda^2 \phi^2$ for
$|\lambda \phi| \ll 1$. In this case, although the amplitude of successive
cycles increases, the increase is much smaller than for a purely
$\lambda^2\phi^2$ potential. This observation supports our conclusion that
hysteresis, and the accompanying increase in the amplitude of successive
expansion maxima, depends crucially upon the ability of the scalar field to
oscillate. The absence of a well-defined minimum in the exponential
potential prevents the field from oscillating, its behaviour is therefore
monotonic and, in the absence of `mixing', the scalar field equation of
state $P/\rho$ does not show any hysteresis. For potentials characteristic
of chaotic inflation, $V \propto \phi^q$, $q > 0$, the extent of
`hysteresis' and the accompanying increase in successive expansion maxima
depend sensitively upon the exponent $q$---reflecting the steepness of the
`chaotic' potential $\phi^q$. For larger $q$, the amount of hysteresis is
smaller, as is the increase in the value of successive maxima of the
expansion factor.

\begin{figure}
\centering \psfig{file=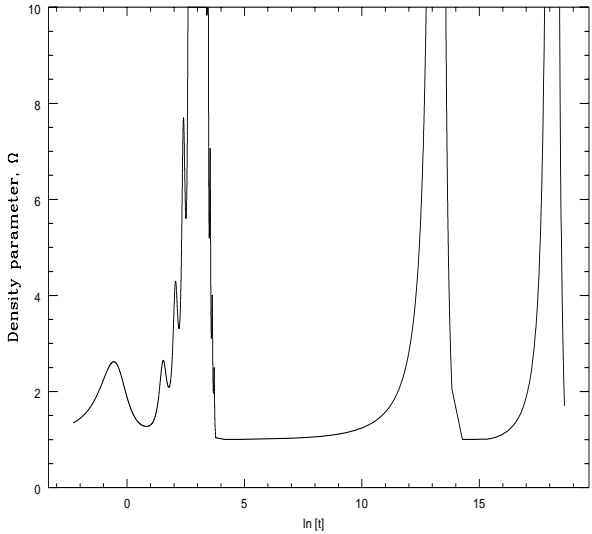,width=3.37truein,height=3.0truein,angle=-90}
\caption{The density parameter $\Omega$ as a function of time ($\Omega \to
\infty$ as the universe recollapses). Note that the time axis is given in
logarithmic units; the oscillations seen (in $\Omega$) in the first cycle
are thus not visible in the later cycles due to the compression of the
scale.  } \label{fig5}
\end{figure}

It is interesting to follow the value of the cosmological density parameter
$\Omega$ as the universe oscillates. From the definition
\begin{equation}
\label{oeqn} \Omega - 1 = (aH)^{-2} \, , \label{eq:om}
\end{equation}
we find  that a larger value of the expansion factor $a$ (at identical $H$)
will result in a value of $\Omega$ which is closer to unity (during
successive cycles). This is borne out by the results of our numerical
analysis shown in Fig.~\ref{fig5}, in which the value of $\Omega$ measured
at identical values of $t - t_{\mathrm min} \sim H^{-1}$ is shown for
successive expansion-contraction cycles [the boundary condition is (ii); the
corresponding expansion factor is shown in Fig.~\ref{fig4a}]. We find that
$\Omega$ approaches unity more closely and for a longer duration during
successive cycles thereby gradually ameliorating the flatness problem.
(During the third expansion cycle, the universe inflates and $\Omega$
approaches unity to great accuracy.)

The time derivative of equation (\ref{oeqn}) yields \\
\begin{equation}
\dot \Omega = H \Omega (1 - \Omega) [ 1 + 3 \omega (t) ] \, , \label{oeqn2}
\end{equation}
where $\omega(t)$ is the general time-dependent equation of state $ \omega =
P(t) / \rho(t)$. Equation (\ref{oeqn2}) shows that the behaviour $\Omega =
{\mathrm const}$ arises if either (i) $\Omega = 1$ or (ii) $\omega(t) \equiv
-1/3$. The value $\Omega = 1$ is the well-known critical point of the
Friedmann equations. The fact that the equation of state $\omega \equiv
-1/3$ also results in a universe evolving with an {\em unchanging value
of\/} $\Omega$ is not so well known, even though this result is completely
general and independent of any assumptions about the curvature of the
universe. It is interesting that oscillations of the scalar field [with
$\omega(t)$ varying between $\omega = + 1$ and $\omega = - 1$] induce
oscillations in $\Omega (t)$, since every time $\omega (t)$ passes through
the critical value $\omega_c = - 1/3$, the value of $\dot \Omega$ changes
sign. Thus, close to the time of recollapse, a large increase in the value
of $\Omega$ is modulated by small oscillations, as shown in Fig.~\ref{fig5}.
The growth in $\Omega$ is due to the increase in the importance of the
curvature term relative to the scalar field density in (\ref{eq:2}), whereas
small modulations in $\Omega$ are due to changes in the equation of state of
the oscillating scalar field.

From (\ref{eq:om}) we find the following simple expression for the total
volume of a closed universe
\begin{equation}
V = {2 \pi^2 \over \left(\Omega_{\mathrm total} - 1\right)^{3/2}}\, V_H \, ,
\label{eq:volume}
\end{equation}
where $V_H = H^{-3}$ is the current Hubble volume.  Observations of
anisotropies in the cosmic microwave background on degree scales appear to
favour a closed universe with $\Omega_{\mathrm total} = 1.11 \pm 0.07$
\cite{boom_max}. By substituting this into (\ref{eq:volume}), we get $258 <
V / V_H < 2467$. The total volume of our universe could therefore be several
thousand times larger than its causal horizon! We also obtain an estimate of
the value of the expansion factor at recollapse ($a_{\mathrm max}$) 
and the corresponding recollapse `redshift' ($z_{\mathrm recoll}$) by
setting to zero the (future) value of the Hubble parameter \cite{ss2000}
\begin{equation}
H(z) = H_0 (1 + z) \left[ 1 - \Omega_{\mathrm total} + \sum_{\alpha}
\Omega_{\alpha} (1 + z)^{\gamma_{\alpha}} \right]^{1/2} \, , \label{eq:age4}
\end{equation}
where $\Omega_\alpha$ is the fraction corresponding to matter of type
$\alpha$ in $\Omega_{\mathrm total}$, $\Omega_{\mathrm total} =
\sum_{\alpha} \Omega_{\alpha}$, $\gamma_\alpha = 1 + 3 w_\alpha$, and $1 + z
= a_0 / a$. Specialising to a CDM dominated universe and assuming for
simpicity that {\em all\/} the matter in the universe is pressureless so
that $\Omega_{\mathrm total} = \Omega_m$, we get
\begin{equation}
H(z) = H_0 (1+z)\left[1 + \Omega_m z\right]^{1/2}
= 0 \, , \label{eq:age6}
\end{equation}
which leads to $z_{\mathrm recoll} = -1/\Omega_m$ or
\begin{equation}
{a_{\mathrm max} \over a_0} = {\Omega_m \over \Omega_m - 1} \, .
\end{equation}
Substituting $\Omega_m = 1.11 \pm 0.07$, we obtain $6.5 \lsim a_{\mathrm
max} / a_0 \lsim 26$, i.e., the size of the universe at recollapse is $\sim
10$ times larger than its present size. 
The corresponding age of the universe at recollapse is
\begin{equation}
t_{\mathrm recoll} = \int_{-1/\Omega_m}^\infty \frac{dz}{H(z)(1+z)} \; ,
\end{equation}
giving
\begin{equation}
t_{\mathrm recoll} = \frac{\pi}{2 H_0} \frac { \sqrt{\Omega_m} }{ \left[ \Omega_m - 1\right]}
\sqrt{ - \left( \frac{\Omega_m}{1-\Omega_m} \right) }  
\end{equation}
i.e. $t_{\mathrm recoll} \sim 400 - 4000$ billion years (for $h=0.5$).

The late-time behaviour of a
$\Lambda_{\mathrm CDM}$ model suggested by recent supernova observations
\cite{lambda} is more complex since, if $\Lambda$ is a constant, the
universe need not recollapse at all but could continue expanding forever
\cite{star99,ss2000}. Quintessence fields which generate a time-dependent
$\Lambda$-term result either in an ever-expanding universe (if $w_Q$ always
remains $\lsim - 1/3$) or in recollapse (if the current acceleration of the
universe is a transient phenomenon).

\section{Discussion and Conclusions.}

We have analysed the behaviour of a massive scalar field in a closed FRW
universe and shown that oscillations of the field about the minimum of its
potential can lead to an asymmetry in its equation of state
$\omega=P_\phi/\rho_\phi$ during the expansion and collapse epochs. This
asymmetry is reflected by the presence of a hysteresis-like feature in
$\omega$ which is, in turn, related to the amount of `work done' by/on the
scalar field during a given expansion-contraction cycle. An important
consequence of this effect is that the imposition of bouncing conditions at
an appropriate early time causes the work done during a given expansion
cycle to be converted into expansion energy, resulting in the growth in
amplitude of each successive expansion maximum in an oscillating universe.
The increase in the value of the expansion maximum results in a recycled
universe which is considerably longer-lived and more flat during each
successive expansion epoch. Thus, the flatness problem is gradually
ameliorated in this model.

Our results also have a direct bearing on the issue of initial conditions in
cosmology. We have shown that even in the worst-case scenario in which the
curvature term dominates, causing the universe to collapse prematurely
without inflating, subsequent cycles will ensure that the value of the
scalar field (or `level of the potential') increases at the commencement of
each expansion cycle, until the value of $\phi$ and $V(\phi)$ become large
enough for inflation to occur. Thus, even if the universe did not inflate
the first time around, it will eventually do so, due to the growing
amplitude of $\phi$ and $V(\phi)$ at the commencement of each new expansion
cycle. Inflation in closed models therefore turns out to be remarkably
robust, provided the universe bounces when it reaches a high density
\cite{kamen}.

We note, finally, that observations of intermediate-angle anisotropies in
the Cosmic Microwave Background made by the BOOMERanG satellite appear to
favour a closed universe, with $\Omega \simeq 1.11 \pm 0.07$ indicated by a
combined analysis of BOOMERANG-98 and MAXIMA-1 data \cite{boom99,boom_max}.
Furthermore, although we have not specified the exact nature or origin for
our scalar field, one might be tempted to view it either as the inflaton or
an inflationary relic. Indeed, the possibility that relic scalar fields
might play the role of dark matter in the universe is very tempting and has
been discussed in \cite{kofman,peebles,sahni_wang}.


\begin{thebibliography}{}
\bibitem[]{byline} Electronic addresses :
\bibitem[\star]{byline} nissim@ncra.tifr.res.in
\bibitem[\dagger]{byline} varun@iucaa.ernet.in
\bibitem[\ddagger]{byline} shtanov@gluk.org
\bibitem{eliade}
M.~Eliade, {\em The Myth of the Eternal Return}, Pantheon, New York (1934).
\bibitem{jaki}
S.~L.~Jaki, {\em Science and Creation: From eternal Cycles to an Oscillating
Universe}, Science History Publ., New York (1974).
\bibitem{munitz}
M.~K.~Munitz, {\em Theories of the Universe}, Free Press, New York (1957).
\bibitem{zimmer}
In Greek mythology the Phoenix---rising from the ashes of a previous
existance---symbolises this idea, while Hindu philosophy narrates the
following description of a cosmic cycle: ``One thousand
mahayugas---4,320,000,000 years of human reckoning---constitute a single day
of Brahma, a single kalpa $\ldots$ \ I have known the dreadful dissolution
of the universe. I have seen all perish again and again, at every cycle. At
that terrible time, every single atom dissolves into the primeval, pure
waters of eternity, whence all originally arose'' $\ldots$ and a new cycle
begins afresh; see, for instance, H.~Zimmer, {\em Myths and Symbols in
Indian Art and Civilization}, Bollingen Foundation, Washington, D.C. (1946);
paperback reprint, Harper and Row, New York (1962);
C.~W.~Misner, K.~S.~Thorne, and J.~A.~Wheeler, {\em Gravitation},
W.~H.~Freeman and Co., NY (1973), p.~752.
\bibitem{tolman}
R.~C.~Tolman, {\em Relativity, Thermodynamics and Cosmology}, Clarendon
Press, Oxford (1934).
\bibitem{zn83}
Ya.~B.~Zeldovich and I.~D.~Novikov, {\em The Structure and Evolution of the
Universe}, University of Chicago Press, Chicago, (1983).
\bibitem{fomin75}
E.~P.~Tryon, Nature, {\bf 246}, 396 (1973); \ 
P.~I.~Fomin,  {\it Gravitational Instability of Vacuum and 
the Cosmological Problem} [in Russian], Preprint ITP-73-137P, 
Institute for Theoretical Physics, Kiev (1973); \ P.~I.~Fomin, 
Dokl.\@ Akad.\@ Nauk Ukrainskoi SSR.\@ Ser.~A, No.~9, 831 (1975).
\bibitem{rees}
M.~J.~Rees, The Observatory {\bf 89}, 193 (1969).
\bibitem{star79}
A.~A.~Starobinsky, JETP Lett.\@ {\bf 30}, 719 (1979)
\bibitem{hoyle1}
F.~Hoyle, G.~Burbidge, and J.~V.~Narlikar, Ap.\@ J.\@ {\bf 410}, 437 (1993).
\bibitem{hoyle2}
F.~Hoyle, G.~Burbidge, and J.~V.~Narlikar, MNRAS {\bf 267}, 1007 (1994).
\bibitem{haw}
S.~W.~Hawking and G.~F.~R.~Ellis, {\em The Large Scale Structure of
Spacetime}, Cambridge University Press, Cambridge (1973).
\bibitem{bd82}
N.~D.~Birrell and P.~C.~W.~Davies, {\em Quantum Fields in Curved Space},
Cambridge University Press, Cambridge (1982).
\bibitem{markov}
M.~A.~Markov, Ann.\@ Phys.\@ (N.Y.\@) {\bf 155}, 333 (1984).
\bibitem{mukhanov}
V.~Mukhanov and R.~Brandenberger, Phys.\@ Rev.\@ Lett.\@ {\bf 68}, 1969
(1992); \ R.~Brandenberger, V.~Mukhanov and A.~Sornborger, Phys.\@ Rev.\@ D
{\bf 48}, 1629 (1993).
\bibitem{linde}
A.~D.~Linde, Phys.\@ Lett.\@ B {\bf 211}, 29 (1988).
\bibitem{barrow}
J.~D.~Barrow and M.~P.~D\c{a}browski, MNRAS {\bf 275}, 850 (1995).
R.~D\"{u}rrer and J.~Laukenmann, Class.\@ Quant.\@ Grav.\@ {\bf 13},
1069 (1996).
\bibitem{kuz}
V.~E.~Kuzmichev and V.~V.~Kuzmichev, Preprints ITP-94-35E, ITP-94-39E, Kiev
(1994); \ V.~V.~Kuzmichev, Yad.\@ Fiz.\@ {\bf 60}, 1707 (1997).
\bibitem{fomin}
P.~I.~Fomin, unpublished.
\bibitem{boom_max}
A.~Jaffe {\em et al.}, astro-ph/0007333.
\bibitem{wein}
S.~Weinberg, {\em Gravitation and Cosmology}, Wiley, New York (1972).
\bibitem{zel}
Ya.~B.~Zeldovich, {\em My Universe: Selected Reviews}, edited by
B.~Ya.~Zeldovich and M.~V.~Sazhin, Harwood Academic (1992).
\bibitem{ll}
L.~D.~Landau and E.~M.~Lifshitz, {\em The Classical Theory of Fields},
Butterworth/Heinemann (1973).
\bibitem{linde90}
A.~D.~Linde, {\em Particle Physics and Inflationary Cosmology}, Harwood
Academic Publishers, Chur (1990).
\bibitem{peebles}
P.~J.~E.~Peebles and A.~Vilenkin, Phys.\@ Rev.\@ D {\bf 59}, 063505 (1999);
P.~J.~E.~Peebles, astro-ph/0002495.
\bibitem{lucchin}
F.~Lucchin and S.~Matarrese, Phys.\@ Rev.\@ D {\bf 32}, 1316 (1985).
\bibitem{ss2000}
V.~Sahni and A.~A.~Starobinsky, Int.\@ J.\@ Mod.\@ Phys.\@ D {\bf 9}, 373
(2000).
\bibitem{sfs92}
V. ~Sahni, H. ~Feldman and A. ~Stebbins, Ap.\@ J.\@ {\bf 385} 1 (1992).
\bibitem{sahni_wang}
V.~Sahni and L.~Wang, Phys.\@ Rev.\@ D (In press), astro-ph/9910097.
\bibitem{lambda}
S.~J.~Perlmutter {\em et al.}, Nature {\bf 391}, 51 (1998); \
S.~J.~Perlmutter {\em et al.}, Appl.\@ Phys.\@ {\bf 517}, 565 (1999); \
A.~G.~Riess {\em et al.}, Astron.\@ J.\@ {\bf 116}, 1009 (1998).
\bibitem{star99}
A.~A.~Starobinsky, astro-ph/9912054.
\bibitem{kamen}
Although the universe was made to re-expand by the imposition of certain
boundary conditions in this paper, a scalar-field dominated closed
cosmological model can also `bounce' intrinsically. The reason for this has
to do with the fact that the scalar field energy momentum tensor does not
always satisfy the energy conditions which require the universe to
recollapse into a singular state. In fact, the dynamics of a scalar-field
dominated closed universes can be quite complex, displaying both chaos and
fractal behaviour, as demonstrated in A.~Yu.~Kamenshchik, I.~M.~Khalatnikov,
and A.~V.~Toporensky, Int.\@ J.\@ Mod.\@ Phys.\@ {\bf D6}, 673 (1997),
gr-qc/9801064; see also B.~S.~Sathyaprakash, P.~Goswami, and K.~P.~Sinha,
Phys.\@ Rev.\@ D {\bf 33}, 2196 (1986).
\bibitem{boom99}
A.~Melchiorri {\em et al.}, astro-ph/9911445; \ A.~E.~Lange {\em et al.},
astro-ph/0005004.
\bibitem{kofman}
L.~A.~Kofman, A.~D.~Linde, and A.~A.~Starobinsky, Phys.\@ Rev.\@ Lett.\@
{\bf 73}, 3195 (1994).

\end{thebibliography}
\end{document}